\documentclass[preprint,amsmath,amssymb,prd,aps]{revtex4-2}

\usepackage{graphicx}
\usepackage{pstricks}
\usepackage{textcomp}
\usepackage{multirow}

\def\feq{&=&} 
\def\fsupset{&\supset&} 

\def\al{\alpha}
\def\be{\beta}
\def\ga{\gamma}
\def\de{\delta}
\def\ep{\epsilon}

\def\et{\eta}

\def\ka{\kappa}

\def\rh{\rho}

\def\si{\sigma}

\def\ph{\phi}

\def\ch{\chi}

\def\om{\omega}
\def\Ga{\Gamma}

\def\Om{\Omega}

\def\kat{{\widetilde \kappa}}

\def\da{^\dagger}

\def\tr{\textrm{tr}}

\def\cM{{\cal M}}

\def\half{{\textstyle{\frac 1 2}}}

\def\quar{{\textstyle{\frac 1 4}}}
\def\sixt{{\textstyle{\frac 1 6}}}

\def\lsim{\mathrel{\rlap{\lower4pt\hbox{\hskip1pt$\sim$}}
    \raise1pt\hbox{$<$}}}
\def\gsim{\mathrel{\rlap{\lower4pt\hbox{\hskip1pt$\sim$}}
    \raise1pt\hbox{$>$}}}

\newcommand{\beq}{\begin{equation}}
\newcommand{\eeq}{\end{equation}}
\newcommand{\bea}{\begin{eqnarray}}
\newcommand{\eea}{\end{eqnarray}}

\newcommand{\nn}{\nonumber}

\def\etal{{\it et al.}}

\def\Re{\hbox{Re}\,}

\def\uk{\breve{k}}

\begin{document}

\title{Probing Lorentz invariance via \hfil \\
diphoton decays of the Higgs boson}

\author{
V.\ Alan Kosteleck\'y$^1$, Connor A. Petway$^1$, and Nathaniel Sherrill$^2$}

\affiliation{
$^1$Physics Department, Indiana University, 
Bloomington, IN 47405, USA}

\affiliation{
$^2$Institut f{\"u}r Theoretische Physik, Leibniz Universit{\"a}t Hannover,
Appelstraße 2, 30167 Hannover, Germany
}

\date{July 2026} 

\begin{abstract}
The prospects for observing small departures from Lorentz invariance 
are studied using effective interactions of the Higgs boson with photons.
A single operator of mass-dimension five 
suffices to describe all leading signatures of Lorentz violation 
in the diphoton decay channel. 
Comparison of the theoretical modified decay rates 
with observed signal strengths 
measured by the ATLAS and CMS Collaborations at the Large Hadron Collider
indicates an estimated attainable sensitivity 
to the corresponding coefficients for Lorentz violation
of order $10^{-4}$~GeV$^{-1}$. 
\end{abstract}

\let\clearpage\relax

\maketitle

\newpage

\section{Introduction}
\label{Introduction}

A central process in the physics of the Higgs boson $h$ 
is the diphoton decay $h\rightarrow \ga\ga$. 
In the Standard Model (SM), 
no tree-level $h\ga\ga$ coupling appears, 
so the decay $h\rightarrow\ga\ga$ proceeds predominantly 
via one-loop diagrams involving a pair of $W$ bosons or top quarks.
Despite the tiny theoretical SM branching ratio BR$_{\rm SM} \simeq 0.2$\%, 
the diphoton channel played a key role in the discovery 
of the Higgs boson~\cite{ATLAS:2012yve, CMS:2012qbp}, 
in part due to the fine mass resolution 
and clean signal reconstruction over a smooth background. 
A standard measure is the signal strength 
\beq
\label{mu}
\mu_{\ga\ga} \equiv 
\sigma\cdot {\rm BR}_{\rm exp}/\sigma\cdot {\rm BR}_{\rm SM},
\eeq 
where $\sigma$ is the total production cross section for the Higgs boson
and ${\rm BR}_{\rm exp}$ is the experimentally observed branching ratio. 
The current \mbox{ATLAS} and CMS Collaboration results 
show good agreement with the SM, 
with $\mu^{\rm ATLAS}_{\ga\ga} = 1.04^{+0.10}_{-0.09}$~\cite{ATLAS:2022tnm} 
and $\mu^{\rm CMS}_{\ga\ga} = 1.12^{+0.09}_{-0.09}$~\cite{CMS:2021kom}. 
Given this sub-10\% precision on a loop-sensitive process, 
$h\rightarrow\ga\ga$ provides an attractive channel 
to search for new physics beyond the SM.

One intriguing prospective type of new physics involves
tiny observable violations of spacetime symmetries, 
including Lorentz invariance,
that could arise in a unified description 
of the fundamental interactions and gravity
such as strings~\cite{strings}.
In the absence of an unambiguous signal, 
a model-independent approach to studying this idea is desirable. 
Adopting the methods of effective field theory~\cite{sw09}
enables a comprehensive treatment of scenarios 
incorporating violations of spacetime symmetries~\cite{kp}.
In Minkowski spacetime,
the general effective field theory of this type,
known as the Standard-Model Extension (SME)~\cite{sme},
extends the SM 
by adding all possible additional terms at the level of the action 
that violate particle Lorentz invariance 
while preserving general coordinate invariance. 
Each additional term is formed as the product
of a Lorentz-violating field operator
contracted with a coefficient governing the size of its effects.
The terms can be classified
according to the operator mass dimension $d$,
with terms of smaller $d$ expected to dominate at lower energy scales.
For the purposes of this work,
we define the minimal SME prior to electroweak symmetry breaking 
as the subset of terms in the SME action
that are of renormalizable dimension $d\leq4$
and that involve local gauge-invariant operators.
This treatment can be further extended to Riemann and other spacetimes
to include General Relativity and its couplings to the SM.
The result provides a general description
of violations of local Lorentz invariance and diffeomorphism invariance
and of deviations from the equivalence principle~\cite{ak04,kl21}.
The coefficients play the role of backgrounds
that modify the inner product and geodesics for the spacetime,
thereby extending its geometry 
to that of a Finsler manifold~\cite{ak04,ak11}.
In the context of fundamental physics
the coefficients are assumed perturbative,
but SME terms with large Lorentz violation are physically realized
in certain semimetals near nodal points in the band structure
where an emergent $(3+1)$-dimensional Lorentz invariance 
is expressed~\cite{klss25}.
Reviews of the framework and its applications can be found, 
for example, in Refs.~\cite{reviews}. 

In recent years,
this general framework has been used 
to analyze many experimental searches for Lorentz violation
across a multitude of physical systems~\cite{tables}.
However,
numerous terms describing a broad spectrum of potential observables 
remain experimentally unexamined to date.
Notable among these are effects in Minkowski spacetime 
involving the strong and electroweak degrees of freedom~\cite{kl19}. 
Collider experiments provide a natural way to search
for many of these signals.
For example,
CPT-odd terms in the quark sector
leads to unique signals for Lorentz violation
in neutral-meson oscillations~\cite{ak98,ek19},
which have led to exceptionally sensitive searches
with collider experiments,
including
the KTeV, FOCUS, and D0 Collaborations
at Fermilab~\cite{ktev01,focus03,d015},
the BaBar Collaboration 
at the Stanford Linear Accelerator Center~\cite{babar08},
the KLOE Collaboration at the Laboratori Nazionali di Frascati~\cite{kloe14},
and the LHCb Collaboration 
at the Large Hadron Collider (LHC)~\cite{lhcb16}.
Recent theoretical progress has enabled 
a subset of CPT-even and CPT-odd quark interactions 
to be incorporated consistently into high-energy hadronic observables%
~\cite{Berger:2015yha,Kostelecky:2016pyx,as19,Kostelecky:2019fse,%
Lunghi:2020hxn, Belyaev:2024chj,Lunghi:2024ctv}, 
leading to further groundbreaking searches 
for quark-sector Lorentz violation with colliders, 
including via the D0 Collaboration~\cite{D0:2012rbu}, 
the ZEUS Collaboration 
at the Deutsche Elektronen-Synchrotron (DESY)~\cite{ZEUS:2022msi}, 
and the Compact Muon Solenoid (CMS) Collaboration 
at the LHC~\cite{CMS:2024rcv}.
Other collider searches in the strong and electroweak sectors of the SME
have also been 
suggested~\cite{lf16,Lunghi:2018uwj,ktk18,Michel:2019tti,ccp20}. 

The existence of unexplored observables for Lorentz violation 
and the experimentally attractive nature 
of the diphoton channel $h\rightarrow \ga\ga$
naturally motivate collider searches in the Higgs-boson sector.
In the SME framework in Minkowski spacetime,
Lorentz-violating operators involving the Higgs boson
appear at mass dimensions $d\geq 3$~\cite{sme}.
However,
none of the minimal-SME terms with $d=3$ or $4$ 
contain couplings directly generating the decay $h\rightarrow \ga\ga$. 
Instead, 
they first contribute as additional vertices 
within the SM one-loop amplitude for the decay. 
Moreover,
indirect constraints on the coefficients for these terms 
inferred from low-energy precision experiments~\cite{tables}
are far more stringent than could be obtained 
from current or foreseeable measurements in the diphoton channel,
so these effects can reasonably be neglected. 
Direct contributions to collider studies of $h\rightarrow \ga\ga$
can therefore arise only from nonminimal Lorentz-violating operators
of dimension $d\geq 5$.

In this work, 
we show that in fact direct couplings to $h\rightarrow \ga\ga$ 
first appear as terms at mass dimension $d=6$
in the unbroken phase of the nonminimal SME.
These terms can be viewed as generalizations 
of ones appearing in the Lorentz-invariant SM effective field theory
(SMEFT)~\cite{Buchmuller:1985jz,Grzadkowski:2010es,Manohar:2006gz}. 
After symmetry breaking, 
they combine into a single Lorentz-violating operator 
of mass dimension $d=5$.
We demonstrate that this operator modifies
the $h\rightarrow\ga\ga$ decay rate 
in ways distinguishable from conventional SM effects,
notably through modulations with sidereal time,
dependence on the geographic location and orientation of the laboratory, 
and novel kinematics.
In particular,
the modulations of the decay rate due to the Earth's rotation
imply that a reanalysis of existing data 
binned as a function of sidereal time 
can reveal signatures otherwise hidden in standard time-averaged analyses. 
The high-precision ATLAS and CMS measurements of $\mu_{\ga\ga}$ 
thus offer a chance 
to probe directly the corresponding coefficients for Lorentz violation 
for the first time.

This paper is organized as follows. 
In Sec.~\ref{Setup}, 
we identify the relevant Lorentz-violating operators 
and deduce the effective diphoton interaction with the Higgs boson
after electroweak symmetry breaking. 
The result is decomposed into rotationally irreducible components 
to facilitate the identification of distinct experimental signatures. 
In Sec.~\ref{Decay rates}, 
we determine the modified $h\rightarrow\ga\ga$ decay rates
at first and second order in the coefficient components. 
Section~\ref{Sidereal variations} discusses experimental signals, 
including sidereal modulations and boost-dependent kinematics, 
and it estimates the sensitivities achievable with current LHC data. 
A summary and outlook form Sec.~\ref{Summary}. 
Natural units with $\hbar = c = 1$ are used throughout.

\section{Effective operators}
\label{Setup}
The set of Lorentz-violating Higgs couplings of mass dimension $d\leq 6$ 
is tabulated in Table XVIII of Ref.~\cite{kl21}. 
Our focus here is on effects triggering the decay of a Higgs boson $h$ 
into two photons ($h\rightarrow \ga\ga$) 
at leading order in Lorentz violation. 
Calculation reveals that this subset is controlled by four terms
involving operators of mass dimension $d=6$,
\bea
{\cal L}^{(6)}_{\rm Higgs} 
\hskip-5pt&\supset\hskip-5pt& 
\tfrac{1}{4}(\uk_{BB\phi}^{(6)})^{\mu \nu \rh \si}
B_{\mu\nu}B_{\rh \si}\phi\da\phi 
\nn \\
&& 
+ \tfrac{1}{4}(\uk_{BW\phi}^{(6)})^{\mu \nu \rh \si}
B_{\mu\nu}\phi\da W_{\rh \si}\phi 
\nn \\
&&
+\tfrac{1}{4}(\uk_{WW\ph,1}^{(6)})^{\mu \nu \rh \si}\phi\da 
W_{\mu\nu}W_{\rh \si} \phi 
\nn \\
&&
+ \tfrac{1}{4}(\uk_{WW\phi,2}^{(6)})^{\mu \nu \rh \si}
\phi\da \tr\left(W_{\mu\nu}W_{\rh \si} \right)\phi 
\label{unbroken}
\eea
in the unbroken phase,
where $\phi$ is the conventional Higgs-boson doublet
and where the U(1) and SU(2) gauge field strengths 
are $B_{\mu\nu}$ and $W_{\mu\nu}$, respectively.
In Minkowski spacetime,
these terms violate Lorentz symmetry but are CPT invariant.
The coefficients for Lorentz violation 
$(\uk_{BB\phi}^{(6)})^{\mu \nu \rh \si}$, 
$(\uk_{BW\phi}^{(6)})^{\mu \nu \rh \si}$, 
$(\uk_{WW\ph,1}^{(6)})^{\mu \nu \rh \si}$, 
and $(\uk_{WW\phi,2}^{(6)})^{\mu \nu \rh \si}$ 
are normalized to mass dimension GeV$^{-2}$. 
Contraction of the product of field strengths 
with a given coefficient produces a quantity 
that is invariant under general coordinate transformations.
The breve diacritic on the coefficients is a compact notation 
representing a series of contractions of coefficients 
with the vierbein, metric, and Levi-Civita tensor~\cite{kl21}. 
At leading order and in Minkowski spacetime,
only two terms are contained in the series
for each coefficient appearing in Eq.~\eqref{unbroken},
$(\uk^{(6)})^{\mu\nu\rho\sigma} = (k^{(6)}_1)^{\mu\nu\rho\sigma} 
+ (k^{(6)}_2)^{\mu\nu}{}_{\alpha\beta}
\epsilon^{\alpha\beta\rho\sigma}$. 
Although these two terms are nominally distinct,
in practice they lead to indistinguishable physics in Minkowski spacetime 
because they have identical symmetries
and contract with the same product of field operators. 
In contrast, 
in curved spacetime the Levi-Civita tensor 
depends on the square root of the metric,
which introduces a physical distinction between the two terms. 
In what follows,
we can therefore drop the breve notation,
$(\uk^{(6)})^{\mu\nu\rho\sigma} \to (k^{(6)})^{\mu\nu\rho\sigma}$,
without loss of generality.

Extracting the couplings to the electromagnetic field is achieved 
by assuming the standard SU(2)$\times$U(1)$\to$U(1) breaking 
of the electroweak symmetry 
and working in the unitary gauge $\phi = (0,r_\phi)/\sqrt{2}$, 
where $r_\phi = v + h$ with vacuum expectation value $v$ 
and Higgs boson $h$. 
We find the diphoton interactions with the Higgs boson 
are controlled by a single term
involving an operator of mass dimension $d=5$,
\beq
{\cal L}^{(5)}_{h\ga\ga} = 
\tfrac{1}{4} (k_{FFh}^{(5)})^{\mu\nu\rh\si} h F_{\mu\nu}F_{\rh\si}, 
\label{LFFh}
\eeq
where $F_{\mu\nu}$ is the electromagnetic field strength. 
The coefficient $(k_{FFh}^{(5)})^{\mu\nu\rh\si}$ for Lorentz violation 
has dimensions of GeV$^{-1}$ in natural units
and is controlled by the vacuum expectation value $v$, 
by the electroweak mixing angle $\theta_W$, 
and by the four unbroken-phase coefficients introduced in Eq.~\eqref{unbroken}. 
In principle,
other Lorentz-violating effects can induce shifts 
in conventional electroweak parameters~\cite{sme}. 
However, 
in the present context such contributions only appear 
as higher-order terms mixing $(k_{FFh}^{(5)})^{\mu\nu\rh\si}$ 
and the unbroken-phase coefficients. 
Neglecting these subleading contributions, we find
\bea
(k_{FFh}^{(5)})^{\mu\nu\rh\si} 
\feq
v\cos^2\theta_W(k_{BB\phi}^{(6)})^{\mu\nu\rh\si} 
\nn \\
&& 
-\tfrac{1}{4}v\sin2\theta_W(k_{BW\phi}^{(6)})^{\mu\nu\rh\si} 
\nn \\
&&
+\tfrac{1}{4}v\sin^2\theta_W(k_{WW\phi,1}^{(6)})^{\mu\nu\rh\si} 
\nn \\
&& 
+\tfrac{1}{2}v\sin^2\theta_W(k_{WW\phi,2}^{(6)})^{\mu\nu\rh\si} .
\label{kFFh}
\eea

The focus of this work is the direct effects
of the coefficient $(k_{FFh}^{(5)})^{\mu\nu\rh\si}$
on the diphoton decays $h\rightarrow\ga\ga$.
In the analysis to follow,
we take the dominant effects to arise 
either from interference of the 1-loop SM process
with the interaction~\eqref{LFFh}
or from the tree-level process at second order in 
$(k_{FFh}^{(5)})^{\mu\nu\rh\si}$.
In principle,
additional SME coefficients with higher $d$ or from other sectors
could contribute to the analysis,
but it is reasonable to take these to be suppressed or negligible
because terms with higher $d$ are subdominant
and others are experimentally constrained to be tiny
or involve suppressed couplings.

While our focus here is on direct effects in collider experiments,
we note in passing that investigations 
of contributions from indirect processes
to other measurements could also be of interest.
Consider,
for example,
the term 
\beq
{\cal L}^{(4)}_{A} = 
-\tfrac{1}{4} k_F^{\mu\nu\rh\si} F_{\mu\nu}F_{\rh\si},
\label{photon}
\eeq
which appears in the broken phase of the minimal SME
and acts to modify the photon propagator. 
When the Higgs field acquires its expectation value,
the unbroken-phase operators in ${\cal L}^{(6)}_{\rm Higgs}$
given in Eq.~\eqref{unbroken} generate additional tree-level contributions
with the same operator structure as those in ${\cal L}^{(4)}_{A}$.
The net effect on a physical photon therefore arises from a combination of
the coefficients appearing in ${\cal L}^{(4)}_{A}$
and ${\cal L}^{(6)}_{\rm Higgs}$.
Adopting various assumptions for the relative sizes of these contributions 
together with the tight existing experimental limits 
on modifications to photon propagation 
could lead to corresponding constraints 
on the coefficients appearing in ${\cal L}^{(6)}_{\rm Higgs}$
and hence provide information about the Lorentz-violating couplings
${\cal L}^{(5)}_{h\ga\ga}$
of the Higgs boson to the photon.
Constraints of this type are evidently model dependent,
a feature that can be intuitively viewed as arising 
because the propagation of a photon
is a physically different process 
from the interaction of the Higgs boson with two photons.
In a general theoretical framework,
two independent measurements are therefore required
to establish direct constraints on both.

As another example of an indirect constraint,
consider radiative effects arising from the interaction~\eqref{LFFh}.
One involving the coefficient 
$(k_{FFh}^{(5)})^{\mu\nu\rh\si}$ 
is a second-order contribution
to the one-loop photon vacuum polarization,
which generates corrections to the photon propagator 
with the same operator structure 
as the minimal-SME term ${\cal L}^{(4)}_{A}$.
This again suggests that under suitable theoretical assumptions
indirect constraints on $(k_{FFh}^{(5)})^{\mu\nu\rh\si}$ 
could be inferred from known experimental bounds 
on modifications to photon propagation.
The indirect approach in this case necessarily comes with model dependence
not only because cancellations may occur among various contributions
to the modified photon propagator
but also because the interaction~\eqref{LFFh} is nonrenormalizable.
In contrast to these indirect examples, 
the effects on diphoton decays of the Higgs boson
studied here are model independent
in the sense that the interaction~\eqref{LFFh} 
is the term at lowest mass dimension 
that generates direct $h\rightarrow\ga\ga$ couplings, 
so its contributions to the decay rate 
are unambiguous predictions at leading order.

For our analysis,
it is convenient to decompose $(k_{FFh}^{(5)})^{\mu\nu\rh\si}$ 
into components that are irreducible under rotations. 
The procedure is similar to that performed 
in Ref.~\cite{ km02} for the minimal-SME coefficient 
$(k_F)^{\mu\nu\rh\si}$,
which exhibits the symmetries of the Riemann tensor 
with a vanishing double trace
and thus contains 19 independent components. 
The coefficient $(k_{FFh}^{(5)})^{\mu\nu\rh\si}$ 
has an analogous symmetry but contains a total of 21 independent components.
The two additional degrees of freedom govern Lorentz scalars. 
One arises from the dependence of the double trace of 
$hF_{\mu\nu}F_{\rh\si}$ on $h$,
which precludes removing the double trace 
from the theory via a field redefinition. 
The second Lorentz scalar arises 
from the dependence of the totally antisymmetric piece of 
$hF_{\mu\nu}F_{\rh\si}$ on $h$,
which precludes interpreting this operator component
as a surface term in the action. 
The 21 independent components of $(k_{FFh}^{(5)})^{\mu\nu\rh\si}$ 
admit a decomposition as
\bea
(\ka_{DEh})^{jk} 
\feq
-2(k^{(5)}_{FFh})^{0j0k} ,
\nn \\
(\ka_{HBh})^{jk}
\feq
\half \ep^{jpq}\ep^{krs}(k^{(5)}_{FFh})^{pqrs} ,
\nn \\
(\ka_{DBh})^{jk} 
\feq
-(\ka_{HEh})^{kj} = (k^{(5)}_{FFh})^{0jpq}\ep^{kpq}.
\label{kappas}
\eea
Here,
the double trace 
{${(k^{(5)}_{FFh})^{\mu\nu}}{}_{\mu\nu}=\tr[\ka_{DEh}]+\tr[\ka_{HBh}]$ 
and the antisymmetric component 
$\epsilon^{\mu\nu\rho\sigma}{(k^{(5)}_{FFh})_{\mu\nu\rho\sigma}} 
= \tr[\ka_{DBh}] $ 
are generically nonvanishing. 
The subscripts involving ${E}$, ${D}$, ${B}$, ${H}$
are adopted in parallel with the notation
for the analogous decomposition of the minimal-SME coefficient 
$(k_F)^{\mu\nu\rh\si}$ in Ref.~\cite{ km02}.

The matrices $(\ka_{DEh})^{jk}$ and $(\ka_{HBh})^{jk}$ 
mix under rotations, 
while $(\ka_{DBh})^{jk}$ contains two sets of components
that transform independently. 
The rotationally irreducible sets 
governing effects of even ($e$) and odd ($o$) parity are given by
\bea
(\kat_{FFh,e+})^{jk} 
\feq
\half(\ka_{DEh}+\ka_{HBh})^{jk} 
\nn \\ &&
- \sixt\tr[\ka_{DEh}+\ka_{HBh}]\de^{jk},
\nn \\
(\kat_{FFh,e-})^{jk} 
\feq
\half(\ka_{DEh}-\ka_{HBh})^{jk}
\nn \\ &&
-\sixt\tr[\ka_{DEh}-\ka_{HBh}]\de^{jk},
\nn \\
(\kat_{FFh,o+})^{jk} 
\feq
\half(\ka_{DBh}+\ka_{HEh})^{jk}, 
\nn \\
(\kat_{FFh,o-})^{jk} 
\feq
\half(\ka_{DBh}-\ka_{HEh})^{jk}, 
\nn \\
\kat_{FFh,\tr +} 
\feq
\sixt\tr[\ka_{DEh}+\ka_{HBh}],
\nn \\
\kat_{FFh,\tr -} 
\feq
\sixt\tr[\ka_{DEh}-\ka_{HBh}], 
\nn \\
\kat_{\widetilde{F}Fh,\tr -} 
\feq
\sixt\tr[\ka_{DBh}-\ka_{HEh}].
\label{kappatilde}
\eea
The matrices 
$ (\kat_{FFh,e+})^{jk}$, $(\kat_{FFh,e-})^{jk}$, and $ (\kat_{FFh,o-})^{jk}$ 
are symmetric and traceless 
and hence each contain five independent components.
The matrix $(\kat_{FFh,o+})^{jk}$ is antisymmetric
and so involves three independent components. 
The traces $\kat_{FFh,\tr +}$ and $\kat_{\widetilde{F}Fh,\tr -}$ 
govern Lorentz-scalar operators,
while $\kat_{FFh,\tr -}$ controls a rotational pseudoscalar operator.
The decomposition~\eqref{kappatilde} therefore
incorporates all 21 independent components 
of $(k_{FFh}^{(5)})^{\mu\nu\rh\si}$,
as desired.
 
Expressing the Lagrange density~\eqref{LFFh} 
in terms of the irreducible coefficients~\eqref{kappatilde}, 
the electric field $\vec{E}$, and the magnetic field $\vec{B}$
yields 
\bea
{\cal L}^{(5)}_{h\ga\ga} 
\feq
-\half h\big[\kat_{FFh,\tr +}(E^2-B^2) 
+ 2\kat_{\widetilde{F}Fh,\tr -}(\vec{E}\cdot\vec{B}) 
\nn \\
&&
+\kat_{FFh,\tr -}(E^2+B^2)
\nn \\
&&
+\vec{E}\cdot(\kat_{FFh,e+}+\kat_{FFh,e-})\cdot\vec{E}
\nn \\
&&
-\vec{B}\cdot(\kat_{FFh,e+}-\kat_{FFh,e-})\cdot\vec{B}
\nn \\
&&
+2\vec{E}\cdot(\kat_{FFh,o+}+\kat_{FFh,o-})\cdot\vec{B}\big].
\label{LEBh}
\eea
The first two terms involve Lorentz-invariant operators
and appear in the SMEFT~\cite{Manohar:2006gz}.
The third term governs boost violations,
while the remaining three terms describe rotation violations.
This expression directly confirms that
$\kat_{FFh,\tr \pm}$ and $(\kat_{FFh,e\pm})^{jk}$ 
are associated with even-parity effects, 
while odd-parity effects are controlled 
by $\kat_{\widetilde{F}Fh,\tr -}$ and $(\kat_{FFh,o\pm})^{jk}$. 
Note that the coefficients with subscript~$+$ 
distinguish couplings to the electric and magnetic fields 
due to antisymmetry under the interchange 
$\vec{E} \leftrightarrow \vec{B}$, 
whereas those with subscript~$-$ 
are symmetric under this interchange regardless of parity.

\section{$h\rightarrow\ga\ga$ decay rates}
\label{Decay rates}

The one-loop SM amplitude expressed 
in terms of the outgoing photon 4-momenta 
$k_1^\nu = (\omega_1, \vec{k}_1)$ and $k_2^\mu = (\omega_2, \vec{k}_2)$ 
is~\cite{Shifman:1979eb}
\beq
\cM_0 = \frac{\al}{4\pi} (\sqrt{2}G_F)^{1/2} 
F (p_h^2 \eta^{\mu\nu}
-2 {k_1}^\nu {k_2}^\mu)\ep^*_{\mu}(k_1) \ep^*_{\nu}(k_2),
\label{SM1loop}
\eeq
where $\alpha$ and $G_F$ are the fine-structure and Fermi constants, 
$F$ is a dimensionless complex number quantifying the loop contributions, 
$p_h^2 = (k_1+k_2)^2$ is the invariant mass squared of the Higgs boson,
and $\ep^{\mu}(k)$ is the polarization 4-vector 
for the outgoing photon of 4-momentum $k$.
In existing experiments studying the diphoton decays of the Higgs boson,
the electromagnetic calorimeters 
are insensitive to the photon polarizations,
implying a sum over the polarizations must be performed 
in the squared modulus of the amplitude~\eqref{SM1loop}.
The resulting unpolarized differential decay width 
evaluated in the center-of-mass frame is
\beq
\frac{d\Ga_0}{d\Om} = 
\left(\frac{\al}{4\pi}\right)^2 
\frac{G_F|F|^2p_h^4}{16\sqrt{2}\pi^2 E_h},
\label{Gamma0}
\eeq
where $\Ga_0$ is the SM width for the diphoton decay,
$\Omega$ is the photon solid angle,
and $E_h$ is the energy of the $h$.

The SME term ${\cal L}^{(5)}_{h\ga\ga}$ given in Eq.~\eqref{LEBh} 
initiates the diphoton decay $h\rightarrow\ga\ga$ 
starting at tree level. 
Following standard practice, 
we adopt the narrow-width approximation 
to factorize the Higgs-boson production from the decay. 
Tree-level Lorentz violation 
leaves unaffected the Higgs-boson propagator 
and has negligible effect on the integrated width 
$\Gamma_0\approx 4$~MeV~\cite{pdg}, 
implying the narrow-width approximation remains valid. 
The production cross section $\sigma$ is dominated 
by gluon-gluon fusion and other non-electromagnetic channels. 
Under these assumptions, 
the theoretical signal strength is 
\beq
\mu^{\rm th}_{\ga\ga} \approx 1+ d\Gamma/\Gamma_0, 
\label{muth}
\eeq
where $d\Gamma$ is the Lorentz-violating correction. 
In principle,
the comparison of Eq.~\eqref{muth} to Eq.~\eqref{mu} 
then provides access to the 21 independent components 
$(k_{FFh}^{(5)})^{\mu\nu\rh\si}$ for Lorentz violation. 

We begin by calculating the first-order effects 
via interference with the relevant SM amplitude 
${\cal M}_0$ given in Eq.~\eqref{SM1loop}.
The first-order decay rate is proportional to the interference 
$|{\cal M}_{\rm int}|^2 
= {\cal M}_0{\cal M}_1{}^* + {\cal M}_0{}^*{\cal M}_1 
= 2\Re({\cal M}_0{}^*{\cal M}_1)$, 
where we find the Lorentz-violating amplitude ${\cal M}_1$ is given by
\beq
{\cal M}_1 = 
-2 (k_{FFh}^{(5)})^{\mu\nu\rh\si}
k_{1\nu}k_{2\si}\ep^*_{\mu}(k_1) \ep^*_{\rho}(k_2).
\label{M1}
\eeq
The gauge invariance of ${\cal L}^{(5)}_{h\ga\ga}$ 
ensures that ${\cal M}_1$ satisfies the electromagnetic Ward identity 
via substitution of a polarization vector $\epsilon^*_\alpha (k)$ 
with the corresponding four-momentum $k_\alpha$,
as expected.
After summing over polarizations, 
we find
\beq
|{\cal \overline M}_{\rm int}|^2 \propto 
(p_h^2\eta_{\mu\rh}-2{k_1}_\rh {k_2}_\mu)
{k_1}_\nu {k_2}_\si (k_{FFh}^{(5)})^{\mu\nu\rh\si}.
\label{M0M1}
\eeq

Only a subset of the components of $(k_{FFh}^{(5)})^{\mu\nu\rh\si}$ 
generates nonzero contributions to the spin-averaged interference
$|{\cal \overline M}_{\rm int}|^2$.
The nine Ricci-like degrees of freedom in the coefficients 
$(\kat_{FFh,e-})^{jk}$, $(\kat_{FFh,o+})^{jk}$, and $\kat_{FFh,\tr -}$ 
can be expressed covariantly as
\bea
(k_{FFh}^{(5)})^{\mu\nu\rh\si} 
\feq
\half\big(\eta^{\mu\rh}(k_{FFh}^{(5)})^{\nu\si}
-\eta^{\mu\si}(k_{FFh}^{(5)})^{\nu\rh}
\nn \\
&&
-\eta^{\nu\rho}(k_{FFh}^{(5)})^{\mu\si}
+\eta^{\nu\si}(k_{FFh}^{(5)})^{\mu\rh}\big),
\label{nbcoeffs}
\eea
where 
$(k_{FFh}^{(5)})^{\nu\si}=
(k_{FFh}^{(5)})^{\al\nu\hphantom{\mu}\sigma}_{\hphantom{\al\nu}\al}$ 
is symmetric and traceless. 
Assuming the form of Eq.~\eqref{nbcoeffs} in Eq.~\eqref{M0M1} 
yields a vanishing contraction. 
This can be viewed as a consequence 
of the zero spin of the initial Higgs boson
together with conservation of angular momentum for the photons,
which follows from the absence of photon birefringence 
in the combination~\eqref{nbcoeffs}~\cite{km09}.
The contribution from the totally antisymmetric component 
of $(k_{FFh}^{(5)})^{\mu\nu\rh\si}$
proportional to $\kat_{\widetilde{F}Fh,\tr -}$
also vanishes by symmetry, 
since the associated operator is odd under a CP transformation 
while the unpolarized diphoton decay $h\rightarrow\gamma\gamma$ 
is a CP-even process~\cite{Manohar:2006gz}. 
Nontrivial contributions to the spin-averaged interference
$|{\cal \overline M}_{\rm int}|^2$
therefore arise only from the ten components of the coefficients 
$(\kat_{FFh,e+})^{jk}$ and $(\kat_{FFh,o-})^{jk}$ 
that govern rotation violations, 
along with the scalar coefficient $\kat_{FFh,\tr +}$.

After some calculation,
we find that the explicit first-order rate
can be written in the form
\bea
\frac{d\Ga_1}{d\Om} 
\feq
\frac{\al(\sqrt{2}G_F)^{\frac{1}{2}}\Re[F]}{32\pi^3 E_h}
~\big[
-\tfrac{1}{4}p_h^4 \kat_{FFh,\tr +}
\nn \\
&&
+\vec{v}_-\cdot \kat_{FFh,e+}\cdot \vec{v}_-
+\half p_h^2 \vec{k}_1 \cdot\kat_{FFh,e+}\cdot \vec{k}_2
\nn \\
&&
+(\vec{k}_1\times\vec{k}_2)\cdot \kat_{FFh,o-} \cdot\vec{v}_-
\big], 
\label{Gamma1}
\eea
where for compactness here and below we define the combinations of vectors 
$\vec{v}_\pm= \om_2 \vec{k}_1\pm\om_1 \vec{k}_2$.
The symmetry under the interchange 
$\vec{k}_1\leftrightarrow \vec{k}_2$ 
implies the solid angle $\Omega$ can be taken 
with respect to either of the photons. 
Note that the violation of rotation invariance 
introduces a nontrivial angular dependence in the final state. 
A realistic comparison with experiment 
therefore necessitates integration over the detector fiducial volume.

The size of the first-order correction $d\Gamma_1$ 
is determined by the product of the coefficient for Lorentz violation 
with the SM-interference coupling 
$\al\sqrt{G_F}\Re[F] \simeq 10^{-4}$~GeV$^{-1}$.
In contrast,
the second-order correction is determined by the product
of two coefficients for Lorentz violation,
with the SM-interference coupling absent.
This suggest that an experiment with weaker sensitivity 
$\gsim 10^{-4}$~GeV$^{-1}$
to the components of $(k_{FFh}^{(5)})^{\mu\nu\rh\si}$ 
is best suited to extract constraints 
from terms contributing at second order in the coefficient
$(k_{FFh}^{(5)})^{\mu\nu\rh\si}$,
rather than from the first-order result $d\Gamma_1$.
Moreover, 
the form of the amplitude ${\cal M}_1$ 
for second-order effects implies sensitivity 
to all 21 components of $(k_{FFh}^{(5)})^{\mu\nu\rh\si}$.
Taken together, 
these features motivate the explicit calculation
of the decay rate for $h\rightarrow \gamma\gamma$
at second order in Lorentz violation.

The relevant squared amplitude can be obtained directly 
from Eq.~\eqref{M1} by summing over polarizations, 
\beq
|{\cal \overline M}_1|^2 = 
4 (k_{FFh}^{(5)})^{\mu\rh\nu\si}(k_{FFh}^{(5)})_{\mu\al\nu\be} 
k_{1\rh}k_{2\si}k^\al_{1}k^\be_{2}.
\label{M2}
\eeq
Some calculation yields the second-order rate ${d\Ga_2}/{d\Om}$.
It contains the terms
\bea
\frac{d\Ga_2}{d\Om} 
\fsupset
\frac{1}{32\pi^2 E_h}
\big[\quar p_h^4(\kat_{FFh,\tr +}^2 + \kat_{\widetilde{F}Fh,\tr -}^2)
\nn \\
&&
+(\om_1 \om_2+\vec{k}_1\cdot\vec{k}_2)^2\kat_{FFh,\tr -}^2
\nn \\
&&
+\om_1\om_2(\om_1\om_2+\vec{k}_1\cdot\vec{k}_2)\tr[(\kat_{FFh,e+})^2]
\nn \\
&&
-\vec{v}_+\cdot(\kat_{FFh,e+})^2\cdot\vec{v}_+ 
+ (\vec{k}_1\cdot \kat_{FFh,e+}\cdot \vec{k}_2)^2
\nn \\
&&
+ \om_1\om_2(\om_1\om_2-\vec{k}_1\cdot\vec{k}_2)\tr[(\kat_{FFh,e-})^2]
\nn \\
&&
-\vec{v}_-\cdot(\kat_{FFh,e-})^2\cdot\vec{v}_- 
+ (\vec{k}_1\cdot\kat_{FFh,e-}\cdot\vec{k}_2)^2
\nn \\
&&
- \om_1\om_2(\om_1\om_2+\vec{k}_1\cdot\vec{k}_2)\tr[(\kat_{FFh,o+})^2]
\nn \\
&&
+\vec{v}_+\cdot(\kat_{FFh,o+})^2\cdot\vec{v}_+ 
+ (\vec{k}_1\cdot\kat_{FFh,o+}\cdot\vec{k}_2)^2
\nn \\
&&
+\om_1\om_2(\om_1\om_2+\vec{k}_1\cdot\vec{k}_2)\tr[(\kat_{FFh,o-})^2]
\nn \\
&&
-\vec{v}_+\cdot(\kat_{FFh,o-})^2\cdot\vec{v}_+
+ (\vec{k}_1\cdot\kat_{FFh,o-}\cdot\vec{k}_2)^2\big].
\nn \\
\label{Gamma2}
\eea
The complete expression for ${d\Ga_2}/{d\Om}$ is lengthy
and involves also cross terms containing products of distinct coefficients.
The result~\eqref{Gamma2} is relevant for a maximum-reach analysis,
in which limits are obtained on independent coefficient components
taken one at a time.
This procedure is standard practice,
providing a useful search measure of depth and breadth~\cite{tables}.
Note that the rate ${d\Ga_2}/{d\Om}$ depends on the coefficients 
$(\kat_{FFh,e-})^{jk}$, $(\kat_{FFh,o+})^{jk}$, and $\kat_{FFh,\tr -}$
for rotation violation 
and the coefficient $\kat_{FFh,\tr -}$ for boost violation, 
in addition to those appearing in the rate ${d\Ga_1}/{d\Om}$.
Note also that 
both the first- and second-order decay rates~\eqref{Gamma1} and~\eqref{Gamma2}
reduce to the result for the SMEFT~\cite{Manohar:2006gz}
in the special limit of Lorentz invariance.

\section{Signals and sensitivities}
\label{Sidereal variations}

The structure of the first-order rate~\eqref{Gamma1} 
reveals that rotation violations 
governed by the 10 independent components of the coefficients 
$(\kat_{FFh,e+})^{jk}$ and $(\kat_{FFh,o-})^{jk}$ 
control the leading experimental observables 
when the experimental sensitivity  
$\lesssim 10^{-4}$~GeV$^{-1}$. 
For the weaker-sensitivity scenario, 
the second-order rate~\eqref{Gamma2} 
instead determines the leading experimental observables,
which involve both rotation and boost violations
governed by the 19 independent components of the coefficients 
$(\kat_{FFh,e+})^{jk}$, $(\kat_{FFh,o-})^{jk}$,
$(\kat_{FFh,e-})^{jk}$, $(\kat_{FFh,o+})^{jk}$, 
and $\kat_{FFh,\tr -}$.
The remaining coefficients 
$\kat_{FFh,\tr +}$ and $\kat_{\widetilde{F}Fh,\tr -}$
generate Lorentz-invariant shifts 
in the theoretical signal strength $\mu^{\rm th}_{\ga\ga}$.

One striking and unique signature for Lorentz violation
is the appearance of sidereal and annual modulations 
in the measured $h\rightarrow \ga\ga$ rate.
These arise because any Earth-based laboratory is noninertial.
The Earth's rotation about its axis and revolution around the Sun
ensure that the laboratory orientation changes with time,
so the presence of rotation violations
can induce corresponding changes in experimental observables.
The modulations depend on the geographic position and orientation
of the laboratory,
thus implying potentially distinct signatures for each experiment. 

To facilitate direct comparisons between experiments,
it has become standard practice to report 
the results of searches for Lorentz violation
using a canonical cartesian reference frame,
which offers an excellent approximation
to an inertial frame over the relevant time scales
and in which the coefficients for Lorentz violation
can be taken as constant. 
The established choice for this frame is the Sun-centered frame (SCF)
with time coordinate $T$ and spatial coordinates $J\in \{X,Y,Z\}$
defined such that 
the $Z$-axis is parallel to the Earth's rotation axis, 
the $X$-axis points from the Earth to the Sun 
at the 2000 vernal equinox coincident with $T=0$, 
and the $Y$-axis completes a right-handed orthogonal basis~\cite{km02,SCF}. 
It is convenient to introduce also 
a standard right-handed laboratory frame (LF) 
with spatial coordinates $j \in \{x,y,z\}$, 
where the $x$-axis points locally south 
and the $y$-axis locally east,
and a standard right-handed detector frame (DF)
with spatial coordinates $j' \in \{x',y',z'\}$
defined to have $z'$-axis along the beamline direction 
and $x'$-axis horizontal to the Earth's surface
at the location of the laboratory. 

Since the Earth's boost in the SCF is of order $10^{-4}$,
it suffices for our purposes to approximate
the Lorentz transformation connecting the SCF to the LF 
by the rotation 
\beq
R^{jJ} =
\left(\begin{array}{ccc}
\cos\ch \cos\om_\oplus T_\oplus & \cos\ch \sin\om_\oplus T_\oplus & -\sin\ch \\
-\sin\om_\oplus T_\oplus & \cos \om_\oplus T_\oplus & 0 \\
\sin\ch \cos\om_\oplus T_\oplus & \sin\ch \sin\om_\oplus T_\oplus & \cos\ch
\end{array}\right),
\label{SCFLF}
\eeq
where $\chi$ is the LF colatitude
and $\omega_\oplus \simeq 2\pi/T_{\rm sid}$ 
is the sidereal-rotation frequency of the Earth,
with $T_{\rm sid} \simeq$ 23~h 56~min.
The local sidereal time $T_\oplus$ convenient in the LF 
is shifted relative to $T$ by~\cite{Toplus}
\beq
T_\oplus \approx T + \frac{(\lambda - \lambda_0)}{360^\circ}T_{\rm sid} ,
\eeq
where $\lambda$ is the LF longitude and $\lambda_0 \approx 66.25^\circ$.
An additional rotation is required to connect the LF to the DF. 
Assuming an upward-inclined beamline 
with positive angle $\al$ and at an angle $\be$ east of due south, 
the rotation from the LF to the DF is given by
\beq
R^{j' j} =
\left(\begin{array}{ccc}
-\sin\be & \cos\be & 0 \\
-\sin\al\cos\be & -\sin\al\sin\be & \cos\al \\
\cos\al\cos\be & \cos\al\sin\be & \sin\al
\end{array}\right).
\label{LFDF}
\eeq
The total rotation from the SCF to the DF 
is therefore given by the product $R^{j' J}=\sum_j R^{j'j}R^{jJ}$.

The rotations~\eqref{SCFLF} and~\eqref{LFDF} can be used 
to express the noninertial DF coefficients appearing 
in the decay rates~\eqref{Gamma1} and \eqref{Gamma2} 
as linear combinations of the inertial SCF coefficients 
in harmonics of $\omega_\oplus$. 
As an example, 
the DF coefficient $(\kat_{FFh,o+})^{x'y'}$ decomposes 
in terms of SCF coefficients as
\bea
(\kat_{FFh,o+})^{x'y'} 
\feq 
(\kat_{FFh,o+})^{XY}(s_\al c_\ch-c_\al c_\be s_\ch)
\nn \\
&&
+\big[(\kat_{FFh,o+})^{YZ}(s_\al s_\ch +c_\al c_\be c_\ch)
\nn \\
&&
\qquad
-(\kat_{FFh,o+})^{XZ}c_\al s_\be\big]\cos(\om_\oplus T_\oplus)
\nn \\
&&
-\big[(\kat_{FFh,o+})^{XZ}(s_\al s_\ch +c_\al c_\be c_\ch)
\nn \\
&&
\qquad
+(\kat_{FFh,o+})^{YZ}c_\al s_\be\big]\sin(\om_\oplus T_\oplus),
\nn \\
\label{kxy}
\eea
where $s_\al \equiv \sin\al$, $c_\al \equiv \cos\al$,
and where analogous definitions hold for the other angles. 
This implies,
for instance,
that a nonzero component $(\kat_{FFh,o+})^{1'2'}$ 
induces a time-independent offset
controlled by $ (\kat_{FFh,o+})^{XY}$ 
and time-dependent modulations 
governed by $(\kat_{FFh,o+})^{XZ}$ and $(\kat_{FFh,o+})^{YZ}$.

For the coefficient $(\kat_{FFh,o+})^{j'k'}$,
the result \eqref{kxy} contains first harmonics
in the sidereal frequency $\omega_\oplus$,
a result that can be traced to the antisymmetry in $j'k'$.
Expressed in terms of SCF coefficients,
the other three sets of DF coefficients 
$(\kat_{FFh,e+})^{j'k'}$, $(\kat_{FFh,e-})^{j'k'}$, 
and $(\kat_{FFh,o-})^{j'k'}$ 
include harmonics at both $\omega_\oplus$ and $2\omega_\oplus$.  
The first-order rate~\eqref{Gamma1} therefore includes harmonics 
at $\omega_\oplus$ and $2\omega_\oplus$, 
while the second-order rate~\eqref{Gamma2} 
includes harmonics at 
$2\omega_\oplus, 3\omega_\oplus$, and $4\omega_\oplus$.
We can therefore compactly display the time dependence
of both Lorentz-violating decay rates in the generic form 
\beq
\frac{d\Ga_{1,2}}{d\Om}= 
A_0+\sum_{n=1}^{4}
[A_n \cos(n\om_\oplus T_\oplus)+B_n\sin(n\om_\oplus T_\oplus)],
\label{Gharmonics}
\eeq
where the expansion parameter
$A_0$ controls time-independent effects
and parameters $A_n, B_n$ govern the time-dependent ones. 

The time-independent shifts from Lorentz violation parametrized by $A_0$ 
generically depend on the angular LF variable $\chi$ 
and DF variables $\alpha, \beta$,
as manifestly evident in the example~\eqref{kxy}.
In principle,
this feature and the novel kinematical dependence could be leveraged 
in an experimental analysis
to separate the underlying coefficients for Lorentz violation 
from the Lorentz-invariant effects governed by 
$\kat_{FFh,\tr +}$ and $\kat_{\widetilde{F}Fh,\tr -}$. 
A treatment along these lines using real data 
would be of definite interest but lies outside our present scope.
In what follows,
we offer some comments 
on time-dependent signals for rotation violations
determined by the parameters $A_n$ and $B_n$, $n\neq 0$,
and on a kinematical approach to measuring
the coefficient $\kat_{FFh,\tr -}$ for boost violation. 

Consider first rotation violations.
Conventional measurements of $h\rightarrow \gamma\gamma$ typically involve 
binning events as a function of kinematical variables,
independently of the measurement time $T_\oplus$. 
This procedure effectively averages away potential signatures 
for rotation violations involving the parameters $A_n$ and $B_n$, $n\neq 0$.
To probe a dataset for signals from the
18 independent components of the coefficients for rotation violation, 
the data can be binned at the event level in $T_\oplus$.  
Note that a primary advantage of studying sidereal modulations 
is the absence of SM background.
This implies that normalizing the experimentally measured rate 
$\Gamma_{\rm exp}(T_\oplus)$ 
to the leading SM prediction~\eqref{Gamma0} 
defines a time-dependent observable 
reducing to unity in the limit of vanishing Lorentz violation. 
Analogous ideas underpin
the various prior searches and proposals%
~\cite{ak98,ek19,ktev01,focus03,d015,babar08,kloe14,lhcb16,%
Berger:2015yha,Kostelecky:2016pyx,as19,Kostelecky:2019fse,%
Lunghi:2020hxn, Belyaev:2024chj,Lunghi:2024ctv,D0:2012rbu,%
ZEUS:2022msi,CMS:2024rcv,lf16,Lunghi:2018uwj,ktk18,Michel:2019tti,ccp20}
for Lorentz and CPT violation using collider experiments.

While a detailed experimental analysis of sidereal variations 
using a real dataset for diphoton decays $h\rightarrow\ga\ga$
remains to be performed,
we can make reasonable assumptions
to deduce order-of-magnitude sensitivities
to the coefficients for rotation violation 
attainable with existing data.
As an illustrative scenario, 
consider a hypothetical laboratory with colatitude $\ch=50^\circ$, 
beamline orientation $\be=30^\circ$, 
and inclination $\al=1^\circ$. 
Using the second-order decay rate~\eqref{Gamma2} 
with a mean Higgs-boson momentum of 100 GeV, 
integrating the total phase space constrained 
by the pseudorapidity endpoint $|\et|=2.5$,
and assuming that a signal modulation of order $\pm10\%$ is measurable, 
we conservatively estimate that 
a sensitivity of order $10^{-4}$~GeV$^{-1}$ can be attained
to the SCF coefficient $(\kat_{FFh,o+})^{JK}$ for rotation violation.
Given the similarity of the contributions in the decay rate~\eqref{Gamma2}, 
it is reasonable that a generic attainable sensitivity 
to each coefficient for rotation violation 
lies within an order of magnitude of $10^{-4}$ GeV$^{-1}$.
Note that this sensitivity implies that contributions from both
the first- and second-order decay rates~\eqref{Gamma1} and~\eqref{Gamma2}
may be relevant in measurements of the coefficients
$(\kat_{FFh,e+})^{JK}$ and $(\kat_{FFh,o-})^{JK}$.
We remark in passing that extending 
the limit of the phase-space integration 
over the full range of the pseudorapidity $\eta$ 
samples the total theoretical phase space, 
thereby yielding an observable identical in form in the SCF and DF 
and thus independent of time. 
Instead of being viewed as a practical limitation of detector construction, 
the restricted range of the pseudorapidity 
is thus seen to be a required feature 
for these types of searches for Lorentz violation,
establishing an orientation for the detector
that provides access to signals for rotation violations.

Next, we consider boost violations 
governed by the coefficient $\kat_{FFh,\tr -}$.
Setting all other coefficients to zero 
and normalizing to the SM rate~\eqref{Gamma0} yields
\beq
\frac{(d\Ga_2/d\Om)}{(d\Ga_0/d\Om)} =
\frac{16\pi^2}{\sqrt{2}\al^2G_F|F|^2}
\frac{(\om_1 \om_2+\vec{k}_1\cdot\vec{k}_2)^2}{p_h^4}
\kat_{FFh,\tr -}^2.
\label{Gammaboost}
\eeq
Adopting the same experimental and kinematic values 
taken for the illustrative scenario described above
and noting that the momentum dependence generates a factor of order one,
we conservatively estimate that a sensitivity to $\kat_{FFh,\tr -}$
of order $10^{-4}$~GeV$^{-1}$ can be attained. 
Note that the sensitivities 
to both boost- and rotation-violating effects 
could be improved via an event-level analysis 
comparing the invariant-mass distribution $p_h^2$ 
to the product of photon energies $\omega_1 \omega_2$.

\section{Summary and outlook}
\label{Summary}

This work investigates small departures from Lorentz invariance 
in diphoton decays of the Higgs boson.
Starting from the set~\eqref{unbroken}
of Lorentz-violating Higgs couplings at mass dimension $d\leq 6$ 
in the unbroken phase of the SME,
we show that the single operator~\eqref{LFFh} at $d=5$
controls the leading signatures 
of Lorentz violation in $h\rightarrow\ga\ga$. 
The corresponding coefficient $(k_{FFh}^{(5)})^{\mu\nu\rh\si}$ 
contains 21 independent components, 
of which 18 control rotation violations, 
one governs boost violation,
and the remaining two determine Lorentz-invariant effects. 
The coefficient $(k_{FFh}^{(5)})^{\mu\nu\rh\si}$ is decomposed
into the rotationally irreducible sets~\eqref{kappatilde},
which are used to express the first- and second-order 
modified decay rates~\eqref{Gamma1} and~\eqref{Gamma2}, 
respectively. 
At first order in Lorentz violation,
10 independent components for rotation violation 
and one Lorentz-invariant operator contribute 
via interference with the SM one-loop amplitude~\eqref{SM1loop}. 
At second order all 21 components contribute, 
including the nine additional coefficients for rotation and boost violations 
that are absent at first order. 
The second-order contributions dominate over the interference-mediated effects 
when the sensitivity to Lorentz violation 
exceeds the scale $\al\sqrt{G_F}\Re[F] \simeq 10^{-4}$~GeV$^{-1}$.

The modified decay rates exhibit several features
that can be exploited in experimental analyses
searching for Lorentz violation,
including modulations in time
at harmonics of the Earth's sidereal frequency,
dependence on the location of the laboratory
and on the orientation of the detector,
and novel kinematical properties. 
The 18 independent components controlling rotation violations
can be studied by binning existing or future experimental data
for the $h\rightarrow\ga\ga$ rate
at the event level as a function of sidereal time. 
Using a plausible scenario for an experimental geometry and kinematics, 
we estimate sensitivities of order $10^{-4}$~GeV$^{-1}$
can be attained to these components
expressed in the canonical Sun-centered frame. 
We show that the coefficient for boost violation can be accessed 
at a comparable sensitivity
through the dependence of the decay rate~\eqref{Gammaboost}
on the invariant mass. 
An experimental analysis reporting measurements 
of any of these 19 coefficient components 
would constitute the first direct constraint 
on the corresponding Lorentz-violating interactions 
of photons with the Higgs boson.

Future advances in sensitivity to these interactions can be envisaged.
The most direct prospect is the High-Luminosity LHC (HL-LHC) upgrade, 
which is currently expected to begin operations 
in the mid-2030s~\cite{ZurbanoFernandez:2020cco}. 
The HL-LHC is planned to deliver 3000~fb$^{-1}$ 
of integrated luminosity at $\sqrt{s} = 14$~TeV 
to both the ATLAS and CMS detectors, 
totaling about 10 times the combined Run 1--3 datasets~\cite{HiggsHLLHC}. 
Along with the modest increase 
in the Higgs production cross section~\cite{deFlorian:2016spz}, 
this implies a yield of roughly an order of magnitude 
more $h\rightarrow\ga\ga$ events. 
The diphoton coupling to the Higgs boson
is projected to be measured with a precision 
at the few-percent level~\cite{ATL-PHYS-PUB-2025-014,ATL-PHYS-PUB-2025-018}, 
representing a notable improvement 
over the current $\simeq$ 9\% precision on $\mu_{\ga\ga}$. 
This improvement also favors prospective sidereal analyses,
as the sensitivity to rotation violation scales 
with the square root of the number of events excluding the SM background. 
Furthermore, 
neglecting the minute difference in detector colatitudes
and noting the symmetry under exchange of the beamline directions, 
the ATLAS and CMS datasets could be combined
to provide a powerful search for Lorentz violation.
Together, 
these considerations suggest the HL-LHC could attain 
a conservative sensitivity of order $10^{-5}$~GeV$^{-1}$ 
to Lorentz violation in the diphoton channel. 

\section*{Acknowledgments}

This work is supported in part by
the U.S.\ Department of Energy
under grant {DE}-SC0010120,
by the Deutsche Forschungsgemeinschaft
under the Heinz Maier Leibnitz Prize BeyondSM HML-537662082,
and by the Indiana University Center for Spacetime Symmetries.

\end{document}